\documentclass{appolb}
\usepackage{epsfig,amsmath}
\def\B{\beta}
\def\G{\gamma}

\def\nuc#1#2{$^{#1}$#2}

\begin{document}
\title{Superdeformed oblate superheavy nuclei
in the self-consistent approach
\thanks{Presented at Zakopane Conference on Nuclear Physics 2012}%
}
\author{L. Pr\'ochniak$^{a}$
and A.~Staszczak$^{b}$
\address{$^{a}$ Heavy Ion Laboratory, University of Warsaw\\
ul. Pasteura 5a, 02-093 Warszawa\\
$^{b}$Institute of Physics, Maria 
Curie-Sk{\l}odowska University,\\ Pl. M. Curie-Sk{\l}odowskiej 1, 20-031 Lublin, Poland\\
}
}

\maketitle

\begin{abstract} 
The HFB self-consistent method has been applied to study the properties of
several neutron deficient superheavy nuclei with $Z=120-124$, $N=160-168$. 
Their distinctive
feature is the existence of minima of the total HFB energy for strongly
deformed, oblate shapes. The self-consistent results agree quite remarkably
with those currently obtained by using microscopic-macroscopic method.
\end{abstract}

\PACS{{21.60.Jz}, {24.75+i},  {27.90.+b}}
  
\section{Introduction}

Various aspects of physics of superheavy nuclei have been subject to numerous
studies from both experimental and theoretical sides for many years. There
are two main theoretical approaches used to describe and predict the properties
of superheavies: the so called microscopic-macroscopic method and the
self-consistent approach. A recent review of theoretical achievements can be
found e.g. in \cite{x12BE01}. In the present paper we consider exotic, very
neutron-deficient nuclei with $Z=120-124$ and $N=160-168$, which have not
been yet
extensively investigated within the self-consistent approach, in particular
they are not covered by an extensive study \cite{x12ST01}. Our work was
motivated by a paper of Jachimowicz \etal{}\cite{2011JA03} where
some intriguing properties of nuclei from this region were predicted using a
sophisticated micro-macro model. First, most of the considered nuclei have global
minimum of energy for strongly deformed, oblate shapes (we call them
super-oblate, SDO, shapes). It is worth noting that the first hint on the existence of
SDO minima appeared in \cite{1996CW01} within the micro-macro model but was not
confirmed by the self-consistent calculations. Second, the authors of
\cite{2011JA03} pointed out that in spite of the fact that the considered
region lies quite far from the predicted island of stability there are some
mechanisms which could increase half-lives of some of nuclei to a range close
to experimental capabilities. Experimentalists have recently reached $Z=118$
isotopes \cite{2012OG06,2012OG03} and plan to go even further, therefore, studies of
such exotic systems can proceed beyond purely theoretical speculations.

\section{Theoretical method}

Our approach is based on the Hartree-Fock-Bogolyubov theory with a
well-established SkM* variant of the Skyrme effective interaction. This
variant gives reasonable values of barrier heights in transactinide region
which is of particular importance if one wants to study fission properties.
The pairing interaction is a sum of volume and surface $\delta$ force, cf.
formula (15) in \cite{2002DO14}. More details of the theoretical model can
be found in \cite{2009ST14,2012WA26,2011ST07,2010WA10} where this model was
applied to a broad range of heavy and superheavy nuclei. The HFB calculations
have been performed using the HFODD solver, see \cite{x12SC01} and
references therein.

In order to study the dependence of total nuclear energy on deformation we do
not use a parametrization of a family of shapes (as in the
microscopic-macroscopic approach) but we rather perform the HFB 
calculations with constraints on the
values of components of the quadrupole mass tensor:
\begin{eqnarray*}
&&q_{20}
=\langle\Psi|\textstyle \sum^A_{i=1} (3z^2_i-r_i^2)|\Psi\rangle\\
&&q_{22}
=\langle\Psi|\textstyle \sum^A_{i=1} \sqrt{3}(x^2_i-y^2_i)|\Psi\rangle \ \ \
\end{eqnarray*}
These values can be
further  
related to the $\B, \G$ deformation variables:
\begin{eqnarray*}
&&\beta=c\sqrt{q_{20}^2+q_{22}^2}  \\
&&\tan\gamma=q_{22}/q_{20}
\end{eqnarray*}
where $c=\sqrt{\pi/5}/A{R_0^2}$ and 
$R_0^2=3(r_0A^{1/3})^2/5$, $r_0=1.23$~fm.

\section{Results}

Our self-consistent calculations with the Skyrme SkM*
interaction confirm most findings of Jachimowicz \etal{}\cite{2011JA03}.

{\it Potential energy landscapes}.
The HFB total energy of nuclei in the region $Z=120-126$ and $N=160-168$ 
exhibit a well pronounced minimum for $\beta\sim 0.4$, $\gamma=60^{\circ}$
and for several nuclei this is a global minimum. In the case of ellipsoidal
shapes such deformation corresponds to a ratio of axes $\sim 3:2$.

Below in Figs. 1 and 2 we show a sample of results for the $Z=120,122$
isotopes. In the case of $Z=120$ only the $N=166$ isotope has a global
super-oblate minimum but this isotope is particularly interesting because it is
the lightest among SDO nuclei and, moreover, its $\alpha$ decay can be hindered
due to a mechanism proposed in \cite{2011JA03} and described below.

Among the $Z=122$ isotopes the three lightest ones ($N=162-166$) have a big oblate
deformation. Starting from $N=168$ the minimum moves gradually with increasing $N$ to a spherical point. 
Similar evolution can be seen also for $Z=124$ isotopes (not shown in the
paper) but with the jump from an SDO to
normal oblate nucleus at $N=170$.

\def\obrmap#1#2{\includegraphics[scale=#2]{#1.v.sym.eps}
}

\def\scob{0.51}
\begin{figure}[htp]
\obrmap{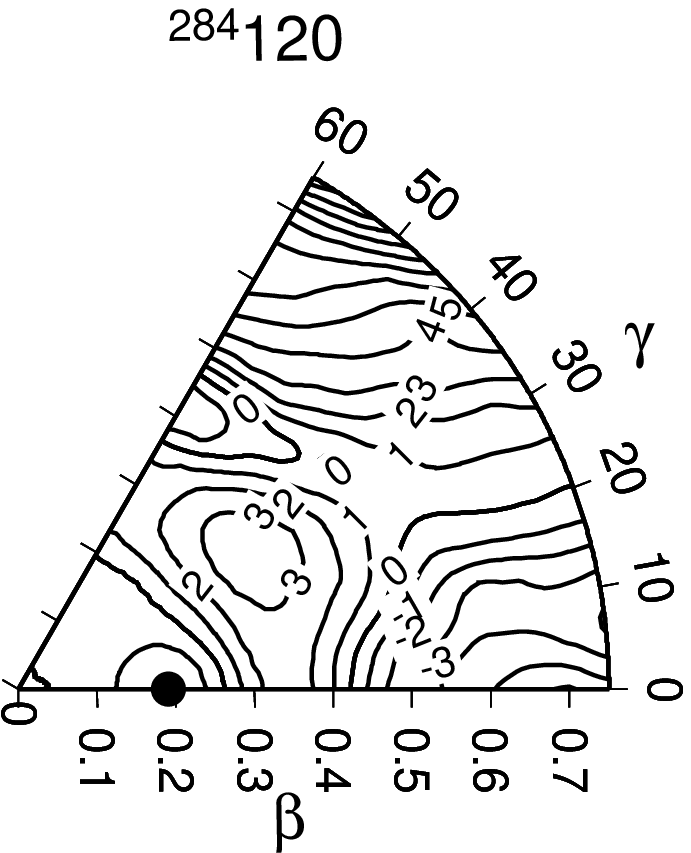}{\scob}\hspace{\fill}\obrmap{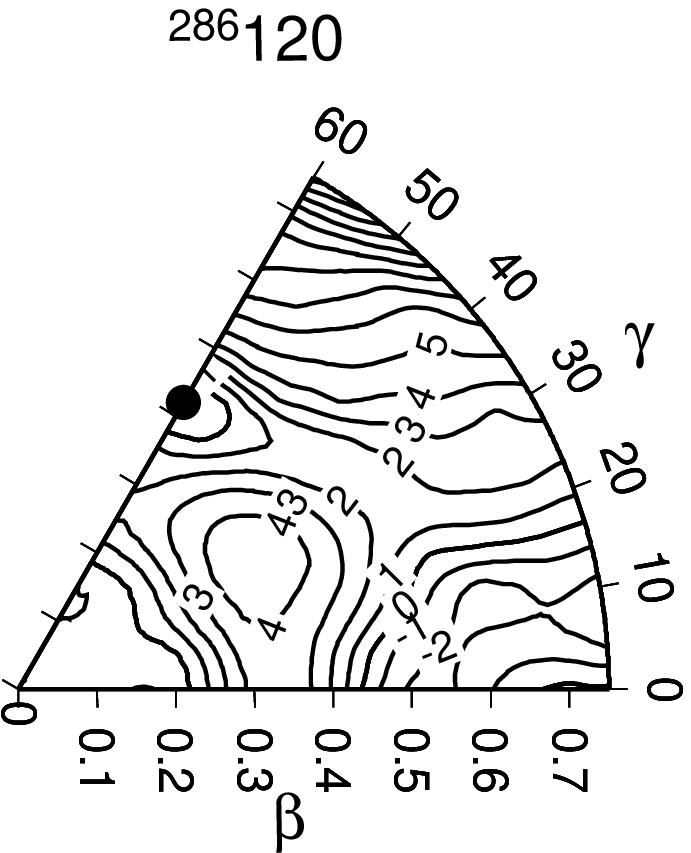}{\scob}\hspace{\fill}\obrmap{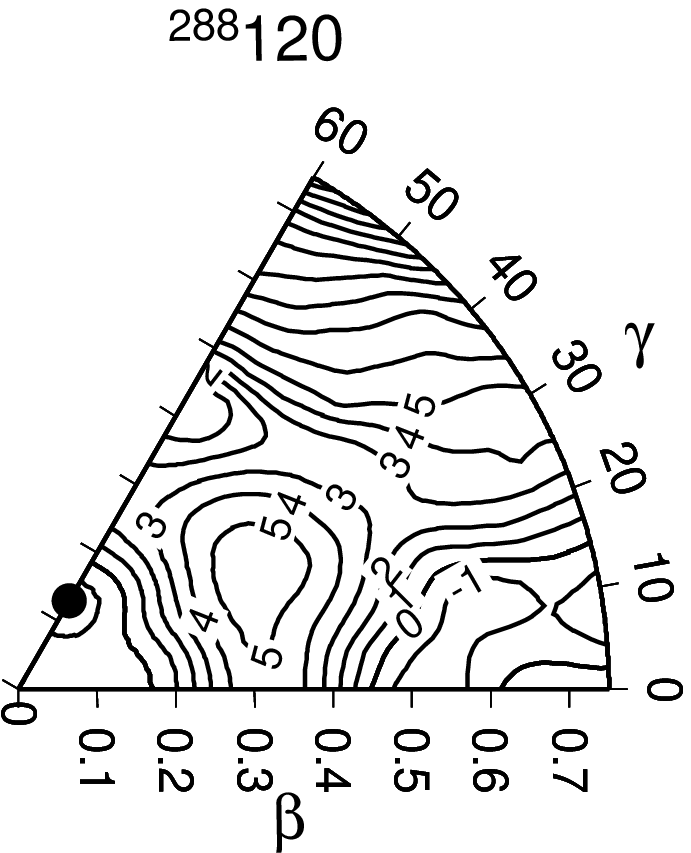}{\scob}
\caption{Plot of the total HFB energy (in MeV, relative to that of a spherical shape) for
the $Z=120$ isotopes.}
\end{figure}

\begin{figure}[htp]
\obrmap{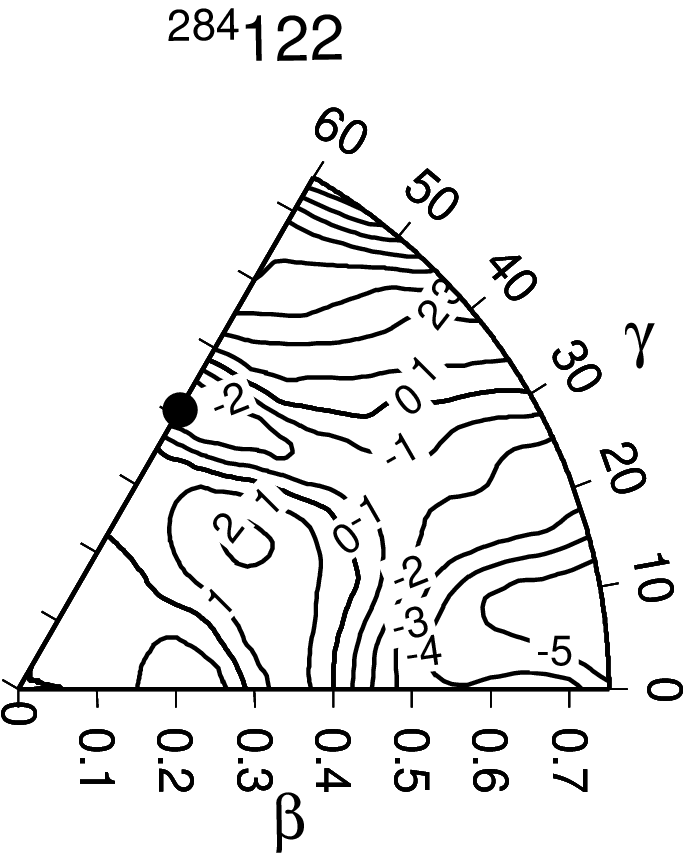}{\scob}\hspace{\fill}\obrmap{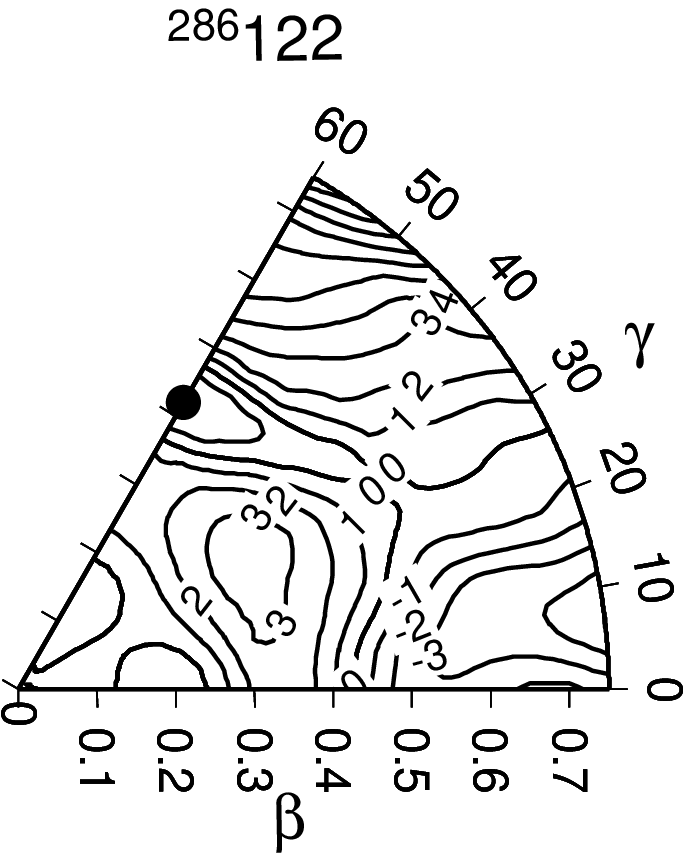}{\scob}\hspace{\fill}\obrmap{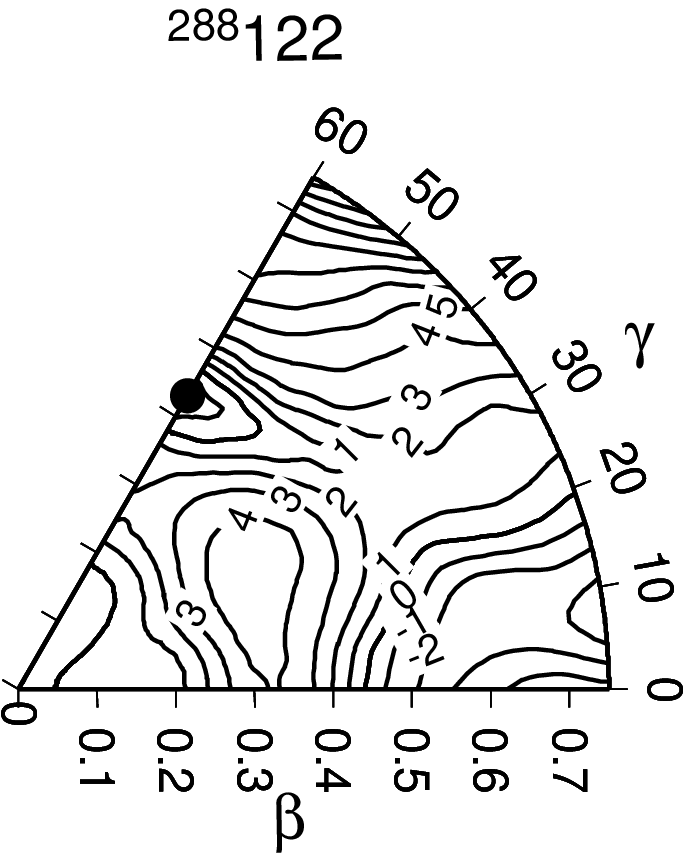}{\scob}

\hfil\obrmap{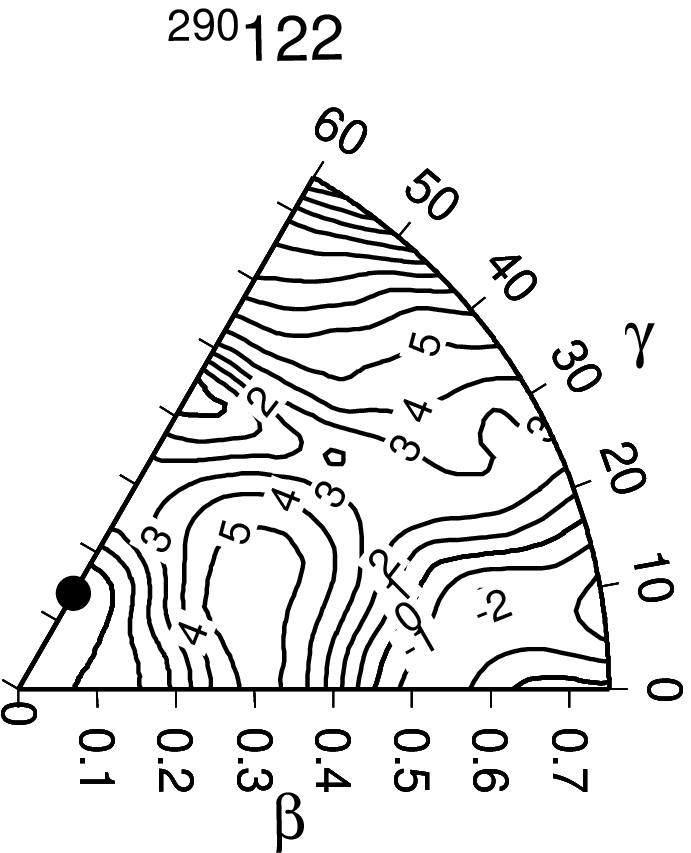}{\scob}\hspace{1cm}\obrmap{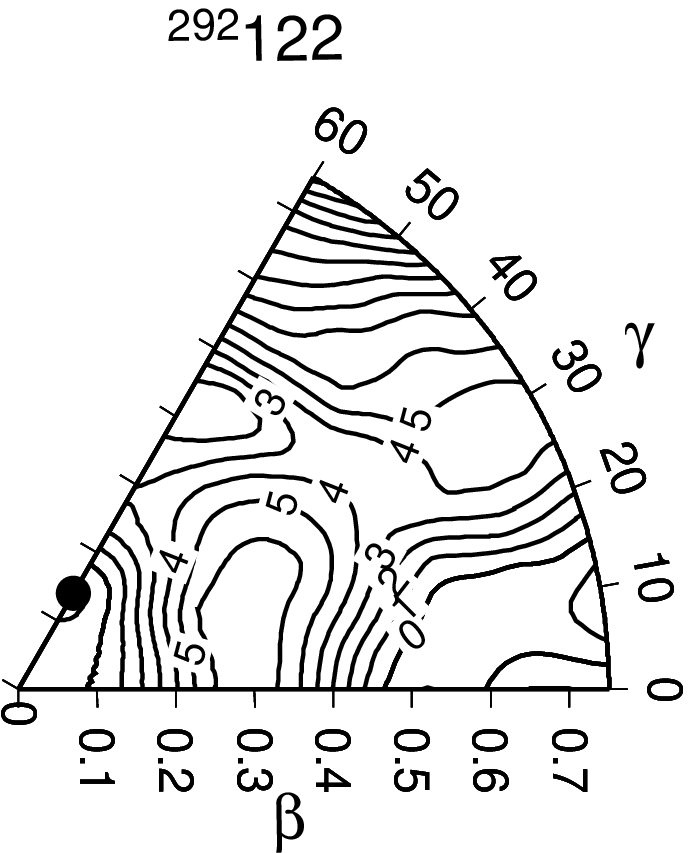}{\scob}\hfil
\caption{Plot of the total HFB energy (in MeV, relative to that of a spherical shape) for
the $Z=122$ isotopes.}

\end{figure}

\begin{figure}[htp]
\includegraphics[scale=0.42]{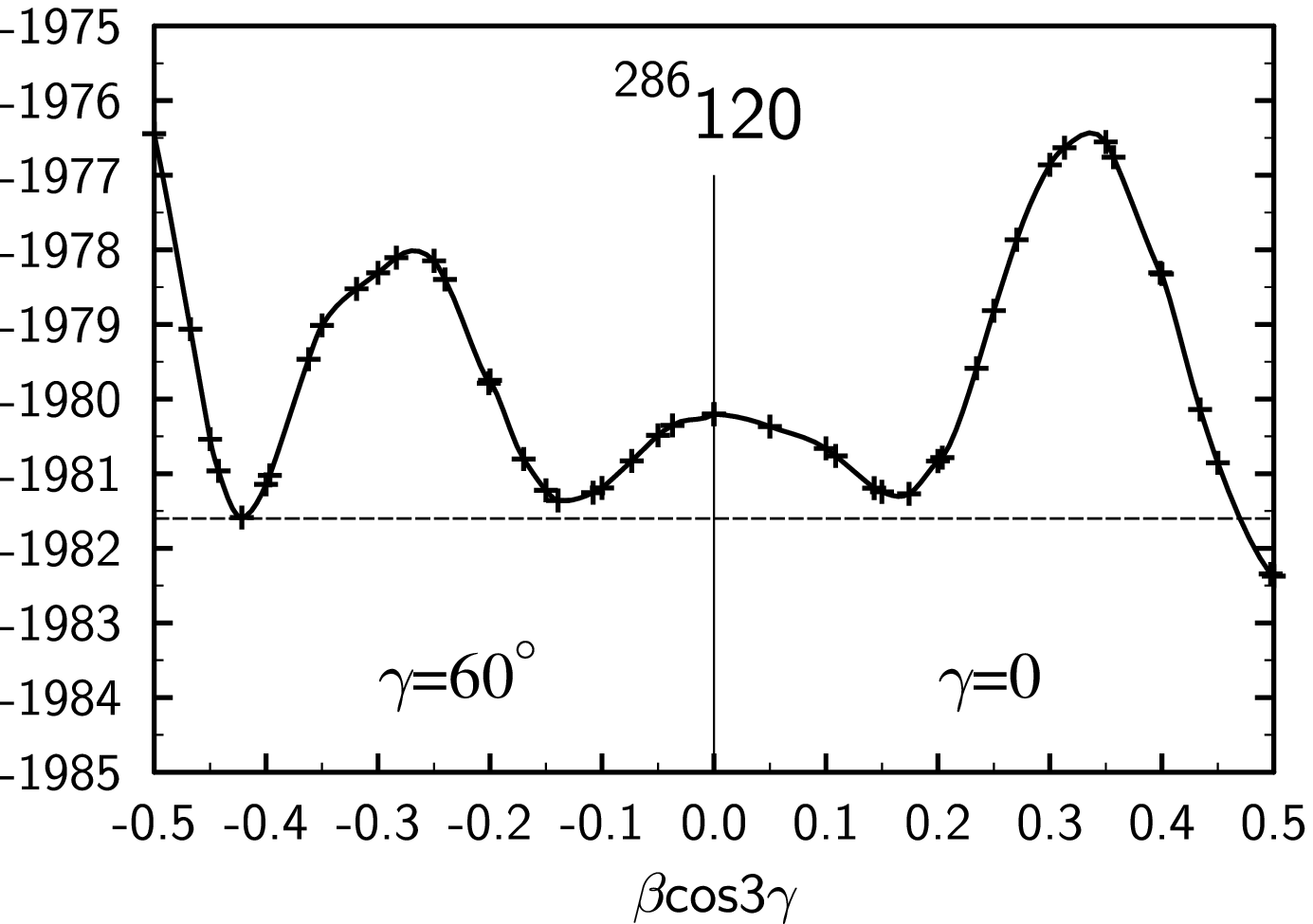}\hspace{5mm} 
\includegraphics[scale=0.42]{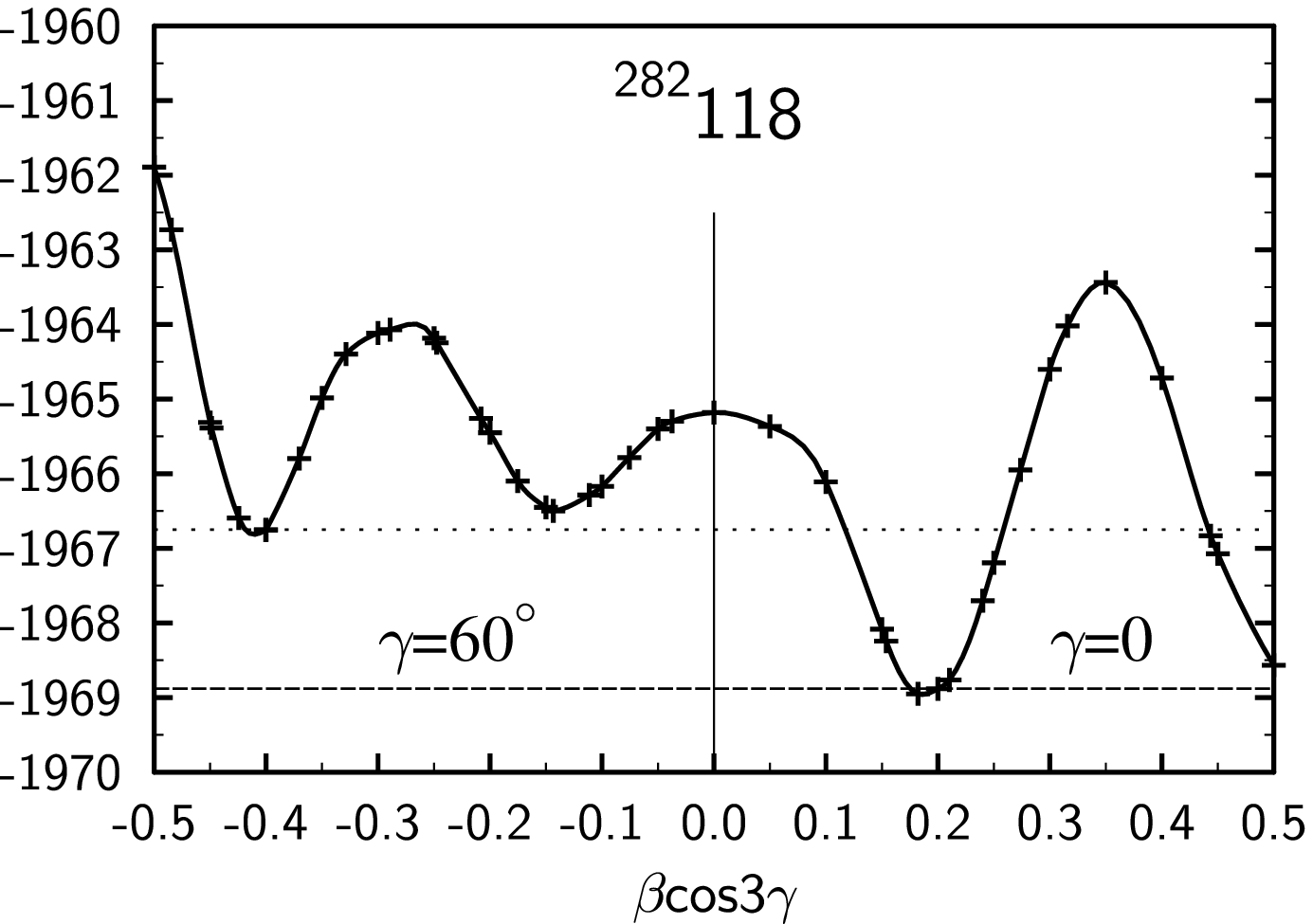}  
\caption{Total energy for axial shapes of \nuc{286}{120} and \nuc{282}{118}
nuclei. In the case of \nuc{282}{118} nucleus prolate and SDO minima
mentioned in the text are
indicated.}
\end{figure}

{\it Decay modes and half-lives estimations.} One should remember that due
to a large deficit of neutrons the considered nuclei are unstable against
proton emission, however, estimations analogous to those in \cite{1996CW01}
give a half-life for this process of an order of days. Much more important
for the existence of considered nuclei are $\alpha$-decay and fission, as can be
seen from systematics of half-lives of isotopes with larger neutron
number e.g. in \cite{x12ST01}. 

{\it $\alpha$-decay.} Typical values of $Q_{\alpha}$ in the
considered region are around 15~MeV which gives $T_{1/2,\alpha}$ around
$10^{-9}$~s (we use here phenomenological formula from \cite{2005PA72}) but
at least in one case (the \nuc{286}{120} nucleus) we could expect a substantial hindrance of the
$\alpha$-decay. As can be seen from Fig.~3 in the daughter nucleus
\nuc{282}{118} there is a (global) prolate but also a super-oblate minimum.
Assuming that $\alpha$ decay from the (super-oblate) \nuc{286}{120} nucleus does not substantially change the internal configuration and goes mainly to a super-oblate shape
of \nuc{282}{118} we obtain the value of $Q_{\alpha}$ for such transition
equal to 13.6 MeV what gives $T_{1/2,\alpha}=10^{-5.84}$~s. 
This mechanism was proposed in \cite{2011JA03} within the micro-macro
approach just for the same pair of nuclei.

{\it Fission.}  As can be seen from Figs.~1,~2 fission process from a
super-oblate minimum goes through triaxial shapes. Within the self-consistent
approach not much is known about
tunneling through such two-dimensional barrier. In particular, one should consider
a full two-dimensional tensor of mass parameters. Work on calculation of
fission half-lives is in progress.

{\it $K$-isomers.} The authors of \cite{2011JA03} pointed out also that long
lived $K$-isomers can appear in both even and odd nuclei in the considered
region. This phenomenon is connected with the existence of single-particle states
with a high angular momentum projection close to the Fermi level. Similar
conclusions can be drawn from our self-consistent calculations. Below we show 
single-particle states around the Fermi level at the minimum of energy for 
\nuc{286}{120}:

\begin{tabular}{ccccc}
\multicolumn{2}{c}{neutrons}& &\multicolumn{2}{c}{protons}\\
$\Omega_{i}$ & {$e_i- e_{F}$ [MeV]}& \rule{15mm}{0mm}&$\Omega_{i}$ &
{$e_i- e_{F}$
[MeV]}  \\
$15/2^{+}$ & $-0.56$ & & $13/2^{+}$ & $-0.66$\\                 
$9/2^{-}$ & $0.75$ & & $7/2^{-}$ & $0.42$
\end{tabular}

\noindent
Here $\Omega_i$ is an eigenvalue of a projection of angular momentum
operator on a symmetry axis for the state $i$.

\section{Conclusions}

The HFB self-consistent calculations have confirmed the predictions of the micro-macro
method concerning the existence of super-oblate ground states for the very
neutron-deficient superheavy nuclei. It is encouraging that both approaches
give very similar results, even on a quantitative level, when they are
extrapolated far from regions where their parameters were fitted.
Some aspects of the discussed subject still need more work, especially
fission properties and odd systems (not touched in the present work).

This work was supported in part by the National
Science Center
(Poland) under Contract DEC-2011/01/B/ST2/03667.

\end{document}